\documentclass[aps,twocolumn,amsmath,amssymb]{revtex4}
\usepackage{graphicx}    \usepackage{rotating}    \usepackage{dcolumn}
\usepackage{epsfig} 
\usepackage{rotating}
\usepackage{amsthm}

\begin{document}

\title{Mutual information rate and bounds for it}

\author{M.  S.  Baptista$^1$, R. M. Rubinger$^2$, E. R. V. Junior$^2$, J. C. Sartorelli$^3$, U. Parlitz$^4$, and C. Grebogi$^{1,5}$} \affiliation{$^1$ Institute for Complex Systems and Mathematical Biology, SUPA, University of Aberdeen, AB24 3UE Aberdeen, United
Kingdom \\
$^2$ Federal University of Itajuba, Av. BPS 1303, Itajub\'a, Brazil\\
$^3$ Institute of Physics, University of S\~ao Paulo, 
Rua do Mat\~ao, Travessa R, 187, 05508-090, S\~ao Paulo, Brazil\\
$^4$ Biomedical Physics Group, 
Max Planck Institute for Dynamics and Self-Organization, 
Am Fassberg 17, 
37077 G\"ottingen,
Germany\\
$^5$ Freiburg Institute for Advanced Studies (FRIAS), University of Freiburg, Albertstr. 19, 79104 Freiburg, Germany}
\date{\today}

\begin{abstract} 
  The amount of information exchanged per unit of time between two
  nodes in a dynamical network or between two data sets is a powerful
  concept for analysing complex systems.  This quantity, known as
  the mutual information rate (MIR), is calculated from the mutual
  information, which is rigorously defined only for random systems.
  Moreover, the definition of mutual information is based on
  probabilities of significant events.  This work offers a simple
  alternative way to calculate the MIR in dynamical (deterministic)
  networks or between two data sets (not fully deterministic), and to
  calculate its upper and lower bounds without having to calculate
  probabilities, but rather in terms of well known and well defined
  quantities in dynamical systems. As possible applications of our
  bounds, we study the relationship between synchronisation and the
  exchange of information in a system of two coupled maps and in
  experimental networks of coupled oscillators.
\end{abstract}

\maketitle 


\section{Introduction}

Shannon's entropy quantifies information \cite{shannon}. It measures
how much uncertainty an observer has about an event being produced by
a random system.  Another important concept in the theory of
information is the mutual information \cite{shannon}.  It measures how
much uncertainty an observer has about an event in a random system
{\bf X} after observing an event in a random system {\bf Y} (or
vice-versa).

Mutual information is an important quantity because it quantifies not
only linear and non-linear interdependencies between two systems or
data sets, but also is a measure of how much information two systems
exchange or two data sets share.  Due to these characteristics, it
became a fundamental quantity to understand the development and
function of the brain \cite{sporns_TCS2004,roland}, to characterise
\cite{juergen_EPJ2009,palus} and model complex systems
\cite{fraser_PRA1986,ulrich_kluwer1998,kantz_book} or chaotic systems,
and to quantify the information capacity of a communication system
\cite{haykin_book}. When constructing a model of a complex system, the
first step is to understand which are the most relevant variables to
describe its behaviour. Mutual information provides a way to identify
those variables \cite{rossi}.

However, the calculation of mutual information in dynamical networks
or data sets faces three main
difficulties\cite{paninski,palus,steuer,papana}.  Mutual information
is rigorously defined for random memoryless processes, only.  In
addition, its calculation involves probabilities of significant events
and a suitable space where probability is calculated. The events need
to be significant in the sense that they contain as much information
about the system as possible. But, defining significant events, for
example the fact that a variable has a value within some particular
interval, is a difficult task because the interval that provides
significant events is not always known. Finally, data sets have finite
size. This prevents one from calculating probabilities correctly. As a
consequence, mutual information can often be calculated with a bias,
only \cite{paninski,palus,steuer,papana}.

In this work, we show how to calculate the amount of information
exchanged per unit of time [Eq. (\ref{MIR_introduction})], the so
called mutual information rate (MIR), between two arbitrary nodes (or
group of nodes) in a dynamical network or between two data sets.  Each
node representing a d-dimensional dynamical system with $d$ state
variables.  The trajectory of the network considering all the nodes in
the full phase space is called ``attractor'' and represented by
$\Sigma$. Then, we propose an alternative method, similar to the ones
proposed in Refs.  \cite{baptista_PRE2008,baptista_PLOSONE2008}, to
calculate significant upper and lower bounds for the MIR in dynamical
networks or between two data sets, in terms of Lyapunov exponents,
expansion rates, and capacity dimension.  These quantities can be
calculated without the use of probabilistic measures. As possible
applications of our bounds calculation, we describe the relationship
between synchronisation and the exchange of information in small
experimental networks of coupled Double-Scroll circuits.

In previous works of Refs.
\cite{baptista_PRE2008,baptista_PLOSONE2008}, we have proposed an
upper bound for the MIR in terms of the positive conditional Lyapunov
exponents of the synchronisation manifold. As a consequence, this
upper bound could only be calculated in special complex networks that
allow the existence of complete synchronisation. In the present work,
the proposed upper bound can be calculated to any system (complex
networks and data sets) that admits the calculation of Lyapunov
exponents.

We assume that an observer can measure only one scalar time series for
each one of two chosen nodes. These two time series are denoted by $X$
and $Y$ and they form a bidimensional set $\Sigma_{\Omega} = (X,Y)$, a
projection of the ``attractor'' into a bidimensional space denoted by
$\Omega$.  To calculate the MIR in higher-dimensional projections
$\Omega$, see Supplementary Information.  Assume that the
space $\Omega$ is coarse-grained in a square grid of $N^2$ boxes with
equal sides $\epsilon$, so $N=1/\epsilon$.

Mutual information is defined in the following way \cite{shannon}.
Given two random variables, {\bf X} and {\bf Y}, each one produces
events $i$ and $j$ with probabilities $P_X(i)$ and $P_Y(j)$,
respectively, the joint probability between these events is
represented by $P_{XY}(i,j)$.  Then, mutual information is defined as
{\small
  \begin{equation}
I_S = H_{X} + H_{Y} - H_{XY}.
\label{IS}
\end{equation}
\noindent 
$H_{X}$ = $-\sum_i P_X(i) \log{[P_X(i)]}$, $H_{Y}$ = $-\sum_j P_Y(j)
\log{[P_Y(j)]}$, and $H_{XY}=-\sum_{i,j} P_{XY}(i,j)
\log{[P_{XY}(i,j)]}$}. For simplification in our notation for the
probabilities, we drop the subindexes $_X$, $_Y$, and $_{XY}$, by
making $P_X(i)=P(i)$, $P_Y(j)=P(j)$, and $P_{XY}(i,j)=P(i,j)$. When
using Eq.  (\ref{IS}) to calculate the mutual information between the
dynamical variables $X$ and $Y$, the probabilities appearing in Eq.
(\ref{IS}) are defined such that $P(i)$ is the probability of finding
points in a column $i$ of the grid, $P(j)$ of finding points in the
row $j$ of the grid, and $P(i,j)$ the probability of finding points
where the column $i$ meets the line $j$ of the grid.

The MIR was firstly introduced by Shannon \cite{shannon} as a ``rate
of actual transmission'' \cite{blanc} and later more rigorously
redefined in Refs.  \cite{dobrushin1959,gray1980}.  It represents the
mutual information exchanged between two dynamical variables
(correlated) per unit of time. To simplify the calculation of the MIR,
the two continuous dynamical variables are transformed into two
discrete symbolic sequences $X$ and $Y$.  Then, the MIR is defined by
\begin{equation}
  MIR=\lim_{n \rightarrow \infty} \frac{I_S(n)}{n}, 
\label{original_MIR}
\end{equation} 
\noindent
where $I_S(n)$ represents the usual mutual information between the two
sequences $X$ and $Y$, calculated by considering words of length $n$.

The MIR is a fundamental quantity in science.  Its maximal value gives
the information capacity between any two sources of information (no
need for stationarity, statistical stability, memoryless)
\cite{verdu}. Therefore, alternative approaches for its calculation or
for the calculation of bounds of it are of vital relevance.  Due to
the limit to infinity in Eq. (\ref{original_MIR}) and because it is
defined from probabilities, the MIR is not easy to be calculated
especially if one wants to calculate it from (chaotic) trajectories of
a large complex network or data sets.  The difficulties faced to
estimate the MIR from dynamical systems and networks are similar to
the ones faced in the calculation of the Kolmogorov-Sinai entropy,
$H_{KS}$ \cite{kolmogorov}, (Shannon's entropy per unit of time).
Because of these difficulties, the upper bound for $H_{KS}$ proposed
by Ruelle \cite{ruelle} in terms of the Lyapunov exponents and valid
for smooth dynamical systems ($H_{KS} \leq \sum \lambda^+_i$, where
$\lambda^+_i$ represent all the $i$ positive Lyapunov exponents) or
the Pesin's equality \cite{pesin} ($H_{KS} = \sum \lambda^+_i$) proved
in Ref.  \cite{ledrapier} to be valid for the large class of systems
that possess a SRB measure, became so important in the theory of
dynamical systems. Our upper bound [Eq. (\ref{I_C})] is a result
equivalent to the work of Ruelle.

\section{Main results}

One of the main results of this work (whose derivation can be seen in
Sec. \ref{methods_MIR}) is to show that, in dynamical networks or data
sets with fast decay of correlation, $I_S$ in Eq. (\ref{IS})
represents the amount of mutual information between $X$ and $Y$
produced within a special time interval $T$, where $T$ represents the
time for the dynamical network (or data sets) to lose its memory from
the initial state or the correlation to decay to zero.  Correlation in
this work is not the usual linear correlation, but a non-linear
correlation defined in terms of the evolution of spatial
probabilities, the quantity $C(T)$ in Sec. \ref{mixing}. Therefore,
the mutual information rate (MIR),
between the dynamical variables $X$ and $Y$ (or two data sets) can be
estimated by {\small
\begin{equation}
MIR=\frac{I_S}{T}
\label{MIR_introduction}
\end{equation}}
\noindent
In systems that present sensitivity to initial conditions, e.g. 
chaotic systems, predictions are only possible for times smaller than
this time $T$. This time has other meanings. It is the expected time
necessary for a set of points belonging to an $\epsilon$-square box in
$\Omega$ to spread over $\Sigma_{\Omega}$ and it is of the order of
the shortest Poincar\'e return time for a point to leave a box and
return to it \cite{gao,baptista_PLA2010}.  It can be estimated by
{\small
\begin{equation}
  T \approx \frac{1}{\lambda_1}\log{\left[ \frac{1}{\epsilon} \right]}.
\label{T} 
\end{equation}}
\noindent
where $\lambda_1$ is the largest positive Lyapunov exponent measured
in $\Sigma_{\Omega}$. Chaotic systems present the mixing property (see
Sec. \ref{mixing}), and as a consequence the correlation $C(t)$ always
decays to zero, surely after an infinitely long time. The correlation
of chaotic systems can also decay to zero for sufficiently large but
finite $t=T$ (see Supplementary Information). $T$ can be interpreted
to be the minimum time required for a system to satisfy the conditions
to be considered mixing. Some examples of physical systems that are
proved to be mixing and have exponentially fast decay of correlation
are nonequilibrium steady-state \cite{nonequilibrium}, Lorenz gases
(models of diffusive transport of light particles in a network of
heavier particles) \cite{sinai_1970}, and billiards \cite{young_2001}.
An example of a ``real world'' physical complex system that presents
exponentially fast decay of correlation is plasma turbulence
\cite{baptista_PHYSICAA2001}. We do not expect that data coming from a
``real world'' complex system is rigorously mixing and has an
exponentially fast decay of correlation.  But, we expect that the data
has a sufficiently fast decay of correlation (e.g. stretched
exponential decay or polynomially fast decays), implying that the
system has sufficiently high sensitivity to initial conditions and as
a consequence $C(t) \cong 0$, for a reasonably small and finite time
$t=T$.

The other two main results of our work are presented in Eqs.
(\ref{I_C_intro}) and (\ref{Icl_intro}), whose derivations are presented in Sec.
\ref{methods_bounds}.  The upper bound for the MIR is given by {\small
  \begin{equation} I_C=\lambda_1-\lambda_2=\lambda_1(2-D),
\label{I_C_intro}
\end{equation}
}
\noindent
where $\lambda_1$ and $\lambda_2$ (positive defined) represent the
largest and the second largest Lyapunov exponent measured in
$\Sigma_{\Omega}$, if both exponents are positive. If the $i$-largest
exponent is negative, then we set $\lambda_i = 0$. If the set $\Sigma_{\Omega}$
represents a periodic orbit, $I_C=0$, and therefore there is no
information being exchanged. The quantity $D$ is defined as {\small
\begin{equation}
D=-\frac{\log{(N_C(t=T))}}{\log{(\epsilon)}}, 
\label{d_intro}
\end{equation}}
\noindent 
where $N_C(t=T)$ is the number of boxes that would be covered by
fictitious points at time $T$. At time $t=0$, these fictitious points
are confined in an $\epsilon$-square box. They expand not only
exponentially fast in both directions according to the two positive
Lyapunov exponents, but expand forming a compact set, a set with no
``holes''. At $t=T$, they spread over $\Sigma_{\Omega}$.

The lower bound for the MIR is given by
\begin{equation}
I_C^l=\lambda_1(2-\tilde{D}_0), 
\label{Icl_intro}
\end{equation}
\noindent
where $\tilde{D}_0$ represents the capacity dimension of the set
$\Sigma_{\Omega}$  
{\small \begin{equation}
    \tilde{D}_0={\lim_{\epsilon \rightarrow 0}} \left[
      -\frac{\log{(\tilde{N}_C(\epsilon))}}{\log{(\epsilon)}} \right],
  \label{tildad}
\end{equation}}
\noindent
where $\tilde{N}_C$ represents the number of boxes in $\Omega$ that
are occupied by points of $\Sigma_{\Omega}$.  

$D$ is defined in a way similar to the capacity dimension, thought it
is not the capacity dimension. In fact, $D \leq \tilde{D}_0$, because
$\tilde{D}_0$ measures the change in the number of occupied boxes in
$\Omega$ as the space resolution varies, whereas $D$ measures the
relative number of boxes with a certain fixed resolution $\epsilon$
that would be occupied by the fictitious points (in $\Omega$) after
being iterated for a time $T$.  As a consequence, the empty space in
$\Omega$ that is not occupied by $\Sigma_{\Omega}$ does not contribute
to the calculation of $\tilde{D}_0$, whereas it contributes to the
calculation of the quantity $D$. In addition, $N_C \geq \tilde{N}_C$
(for any $\epsilon$), because while the fictitious points form a
compact set expanding with the same ratio as the one for which the
real points expand (ratio provided by the Lyapunov exponents), the
real set of points $\Sigma_{\Omega}$ might not occupy many boxes.

\section{Methods}\label{methods}

\subsection{Mixing, correlation decay and  invariant measures}\label{mixing}

Denote by $F^{T}(x)$ a mixing transformation that represents how a
point $x \in \Sigma_{\Omega}$ is mapped after a time $T$ into
$\Sigma_{\Omega}$, and let $\rho(x)$ to represent the probability of
finding a point of $\Sigma_{\Omega}$ in $x$ (natural invariant
density). Let $I^{\prime}_1$ represent a region in $\Omega$. Then,
$\mu(I^{\prime}_1)=\int \rho(x) dx$, for $x \in I^{\prime}_1$
represents the probability measure of the region $I^{\prime}_1$.
Given two square boxes $I^{\prime}_1 \in \Omega$ and $I^{\prime}_2 \in
\Omega$, if $F^{T}$ is a mixing transformation, then for a
sufficiently large $T$, we have that the correlation
$C(T)=\mu[F^{-T}(I^{\prime}_1) \cap I^{\prime}_2] -
\mu[I^{\prime}_1]\mu[I^{\prime}_2]$, decays to zero, the probability
of having a point in $I^{\prime}_1$ that is mapped to $I^{\prime}_2$
is equal to the probability of being in $I^{\prime}_1$ times the
probability of being in $I^{\prime}_2$.  That is typically what
happens in random processes.

If the measure $\mu(\Sigma_{\Omega})$ is invariant, then
$\mu([F^{-T}(\Sigma_{\Omega})]=\mu(\Sigma_{\Omega})$. Mixing and
ergodic systems produce measures that are invariant.

\subsection{Derivation of the mutual information rate (MIR) in
  dynamical networks and data sets}\label{methods_MIR}

We consider that the dynamical networks or data sets to be analysed
present either the mixing property or have fast decay of correlations,
and their probability measure is time invariant. If a system that is
mixing for a time interval $T$ is observed (sampled) once every time
interval $T$, then the probabilities generated by these snapshot
observations behave as if they were independent, and the system
behaves as if it were a random process. This is so because if a system
is mixing for a time interval $T$, then the correlation $C(T)$ decays
to zero for this time interval. For systems that have some decay of
correlation, surely the correlation decays to zero after an infinite
time interval. But, this time interval can also be finite, as shown in
Supplementary Information.

Consider now that we have experimental points and they are sampled
once every time interval $T$. The probability $\tilde{P}_{XY}(i,j)
\rightarrow \tilde{P}_{XY}(k,l)$ of the sampled trajectory to follow a
given itinerary, for example to fall in the box with coordinates
$(i,j)$ and then be iterated to the box $(k,l)$ depends exclusively on
the probabilities of being at the box $(i,j)$, represented by
$\tilde{P}_{XY}(i,j)$, and being at the box $(k,l)$, represented by
$\tilde{P}_{XY}(k,l)$.  Therefore, for the sampled trajectory,
$\tilde{P}_{XY}(i,j) \rightarrow \tilde{P}_{XY}(k,l) =
\tilde{P}_{XY}(i,j)\tilde{P}_{XY}(k,l)$.  Analogously, the probability
$\tilde{P}_{X}(i) \rightarrow \tilde{P}_{Y}(j)$ of the sampled
trajectory to fall in the column (or line) $i$ of the grid and then be
iterated to the column (or line) $j$ is given by $\tilde{P}_{X}(i)
\rightarrow \tilde{P}_{Y}(j) = \tilde{P}_{X}(i)\tilde{P}_{Y}(j)$.

The MIR of the experimental non-sampled trajectory points can be
calculated from the mutual information $\tilde{I}_S$ of the sampled
trajectory points that follow itineraries of length $n$:
\begin{equation}
  MIR=\lim_{n \rightarrow \infty} \frac{\tilde{I}_S(n)}{nT}, 
  \label{original_MIR_sampled}
\end{equation} 

Due to the absence of correlations of the sampled trajectory points,
the mutual information for these points following itineraries of
length $n$ can be written as
\begin{equation}
  \tilde{I}_S(n)=n[\tilde{H}_{X}(n=1) + \tilde{H}_{Y}(n=1) - \tilde{H}_{XY}(n=1)],
\label{original_MIR_sampled1}
\end{equation} 
\noindent
where $\tilde{H}_{X}(n=1)$ = $-\sum_i \tilde{P}_X(i)
\log{[\tilde{P}_X(i)]}$, $\tilde{H}_{Y}(n=1)$ = $-\sum_j
\tilde{P}_Y(j) \log{[\tilde{P}_Y(j)]}$, and
$\tilde{H}_{XY}(n=1)=-\sum_{i,j} \tilde{P}_{XY}(i,j)
\log{[\tilde{P}_{XY}(i,j)]}$, and $\tilde{P}_X(i)$, $\tilde{P}_Y(j)$,
and $\tilde{P}_{XY}(i,j)$ represent the probability of the sampled
trajectory points to fall in the line $i$ of the grid, in the column
$j$ of the grid, and in the box $(i,j)$ of the grid, respectively.

Due to the time invariance of the set $\Sigma_{\Omega}$ assumed to
exist, the probability measure of the non-sampled trajectory is equal
to the probability measure of the sampled trajectory.  If a system
that has a time invariant measure is observed (sampled) once every
time interval $T$, the observed set has the same natural invariant
density and probability measure of the original set. As a consequence,
if $\Sigma_{\Omega}$ has a time invariant measure, the probabilities
$P(i)$, $P(j)$, and $P(i,j)$ (used to calculate $I_S$) are equal to
$\tilde{P}_X(i)$, $\tilde{P}_Y(j)$, and $\tilde{P}_{XY}(i,j)$.


Consequently, $\tilde{H}_{X}(n=1)=H_{X}$, $\tilde{H}_{Y}(n=1)=H_{Y}$,
and $\tilde{H}_{XY}(n=1)=H_{XY}$, and therefore 
$\tilde{I}_S(n) = n I_S(n)$. Substituting into Eq.
(\ref{original_MIR_sampled}), we
finally arrive to
\begin{equation}
  MIR = \frac{I_S}{T} 
\label{MIR}
\end{equation}
\noindent
where $I_S$ between two nodes is calculated from Eq.  (\ref{IS}).

Therefore, in order to calculate the MIR, we need to estimate the time
$T$ for which the correlation of the system approaches zero and the
probabilities $P(i)$, $P(j)$, $P(i,j)$ of the experimental non-sampled
experimental points to fall in the line $i$ of the grid, in the column
$j$ of the grid, and in the box $(i,j)$ of the grid, respectively.

\subsection{Derivation of an upper ($I_C$)  and lower ($I_C^l$) bounds for the MIR}\label{methods_bounds}

Consider that our attractor $\Sigma$ is generated by a 2d expanding
system that possess 2 positive Lyapunov exponents $\lambda_1$ and
$\lambda_2$, with $\lambda_1 \geq \lambda_2$. $\Sigma \in \Omega$.
Imagine a box whose sides are oriented along the orthogonal basis used
to calculate the Lyapunov exponents.  Then, points inside the box
spread out after a time interval $t$ to $\epsilon \sqrt{2}
\exp^{\lambda_1 t}$ along the direction from which $\lambda_1$ is
calculated. At $t=T$, $\epsilon \sqrt{2} \exp^{\lambda_1 T}=L$, which
provides $T$ in Eq.  (\ref{T}), since $L=\sqrt{2}$. These points
spread after a time interval $t$ to $\epsilon \sqrt{2} \exp^{\lambda_2
  t}$ along the direction from which $\lambda_2$ is calculated.  After
an interval of time $t=T$, these points spread out over the set
$\Sigma_{\Omega}$. We require that for $t \leq T$, the distance
between these points only increases: the system is expanding.

Imagine that at $t=T$, fictitious points initially in a square box
occupy an area of $\epsilon \sqrt{2} \exp^{\lambda_2
  T}L=2\epsilon^2\exp^{(\lambda_2+\lambda_1) T}$. Then, the number of
boxes of sides $\epsilon$ that contain fictitious points can be
calculated by $N_C=2\epsilon^2 \exp^{(\lambda_1 +
  \lambda_2)T}/2\epsilon^2=\exp^{(\lambda_1 + \lambda_2)T}$. From Eq.
(\ref{T}), $N=\exp^{\lambda_1T}$, since $N=1/\epsilon$.

We denote with a lower-case format, the probabilities $p(i)$, $p(j)$,
and $p(i,j)$ with which fictitious points occupy the grid in $\Omega$.
If these fictitious points spread uniformly forming a compact set
whose probabilities of finding points in each fictitious box is equal,
then $p(i)=1/N$ ($=\frac{1}{N_C}\frac{N_C}{N}$), $p(j)=1/N$, and
$p(i,j)=1/N_C$.  Let us denote the Shannon's entropy of the
probabilities $p(i,j)$, $p(i)$ and $p(j)$ as $h_X$, $h_Y$, and
$h_{XY}$. The mutual information of the fictitious trajectories after
evolving a time interval $T$ can be calculated by $I_S^u = h_X + h_Y -
h_{XY}$. Since, $p(i)=p(j)=1/N$ and $p(i,j)=1/N_C$, then
$I_S^u=2\log{(N)}-\log{(N_C)}$. At $t=T$, we have that
$N=\exp^{\lambda_1T}$ and $N_C=\exp^{(\lambda_1 + \lambda_2)T}$,
leading us to $I_S^u=(\lambda_1-\lambda_2)T$.  Therefore, defining,
$I_C=I_S^u/T$, we arrive at $I_C=\lambda_1-\lambda_2$.
 
We defining $D$ as {\small
\begin{equation}
D=-\frac{\log{(N_C(t=T))}}{\log{(\epsilon)}}, 
\label{d}
\end{equation}}
\noindent
where $N_C(t=T)$ being  the number of boxes that would be covered by
fictitious points at time $T$. At time $t=0$, these fictitious points
are confined in an $\epsilon$-square box. They expand not only
exponentially fast in both directions according to the two positive
Lyapunov exponents, but expand forming a compact set, a set with no
``holes''. At $t=T$, they spread over $\Sigma_{\Omega}$.

Using $\epsilon=\exp^{-\lambda_1T}$ and $N_C=\exp^{(\lambda_1 +
  \lambda_2)T}$ in Eq.  (\ref{d}), we arrive at
$D=1+\frac{\lambda_2}{\lambda_1}$, and therefore, we can write that
 {\small
  \begin{equation} I_C=\lambda_1-\lambda_2=\lambda_1(2-D),
\label{I_C}
\end{equation}
}

To calculate the maximal possible MIR, of a random independent
process, we assume that the expansion of points is uniform only along
the columns and lines of the grid defined in the space $\Omega$,
i.e., $P(i)=P(j)=1/N$, (which maximises $H_X$ and $H_Y$), and we allow
$P(i,j)$ to be not uniform (minimising $H_{XY}$) for all $i$ and $j$,
then {\small
\begin{equation}
I_S(\epsilon)= -2\log{(\epsilon)}+ \sum_{i,j}P(i,j)\log{[P(i,j)]}.
\label{IS_lower}
\end{equation}}
\noindent
Since $T(\epsilon)=-1/\lambda_1 \log{(\epsilon)}$, dividing
$I_S(\epsilon)$ by $T(\epsilon)$, taking the limit of $\epsilon
\rightarrow 0$, and reminding that the information dimension of the
set $\Sigma_{\Omega}$ in the space $\Omega$ is defined as
$\tilde{D}_1$=$\lim_{\epsilon \rightarrow 0}
\frac{\sum_{i,j}P(i,j)\log{[P(i,j)]}}{\log{(\epsilon)}}$, we obtain
that the MIR is given by {\small
\begin{equation}
I_S/T = \lambda_1(2-\tilde{D}_1).
\label{almost_true1}
\end{equation}}

Since $\tilde{D}_1 \leq \tilde{D}_0$ (for any value of $\epsilon$),
then $\lambda_1(2-\tilde{D}_1) \geq \lambda_1(2-\tilde{D}_0)$, which
means that a lower bound for the maximal MIR [provided by Eq.
(\ref{almost_true1})] is given by {\small 
\begin{equation}
I_C^l=\lambda_1(2-\tilde{D}_0), 
\label{Icl}
\end{equation}}

\noindent
But $D \leq \tilde{D}_0$
(for any value of $\epsilon$), and therefore $I_C$ is an upper bound
for $I_C^l$.

To show why $I_C$ is an upper bound for the maximal possible MIR,
assume that the real points $\Sigma_{\Omega}$ occupy the space
$\Omega$ uniformly.  If $\tilde{N}_C > N$, there are many boxes being
occupied. It is to be expected that the probability of finding a point
in a line or column of the grid is $P(i)=P(j) \cong 1/N$, and $P(i,j)
\cong 1/\tilde{N}_C$. In such a case, $MIR \cong I_C^l$, which implies
that $I_C \geq MIR$.  If $\tilde{N}_C < N$, there are only few boxes
being sparsely occupied.  The probability of finding a point in a line
or column of the grid is $P(i)=P(j) \cong 1/\tilde{N}_C$, and $P(i,j)
\cong 1/\tilde{N}_C$. There are $\tilde{N}_C$ lines and columns being
occupied by points in the grid.  In such a case, $I_S \cong
2\log{(\tilde{N}_C)}-\log{(\tilde{N}_C)} \cong \log{(\tilde{N}_C)}$.
Comparing with $I_S^u=2\log{(N)}-\log{(N_C)}$, and since $\tilde{N}_C
< N$ and $N_C \geq \tilde{N}_C$, then we conclude that $I_S^u \geq
I_S$, which implies that $I_C \geq MIR$.

Notice that if $P(i,j)=p(i,j)=1/N_C$ and $\tilde{D}_1 = \tilde{D}_0$,
then $I_S/T=I_C^l=I_C$.

\subsection{Expansion rates}

In order to extend our approach for the treatment of data sets coming
from networks whose equations of motion are unknown, or for
higher-dimensional networks and complex systems which might be neither
rigorously chaotic nor fully deterministic, or for experimental data
that contains noise and few sampling points, we write our bounds in
terms of expansion rates defined in this work by {\small
\begin{equation}
  e_k(t)=1/\tilde{N}_C \sum_{i=1}^{\tilde{N}_C} \frac{1}{t}  log{[L_k^i(t)]},  
\label{define_exp_rates}
\end{equation}}
\noindent
where we consider $k=1,2$. $L^i_1(t)$ measures the largest growth rate
of nearby points. In practice, it is calculated by
$L^i_1(t)=\frac{\Delta}{\delta}$, with $\delta$ representing the
largest distance between pair of points in an $\epsilon$-square box
$i$ and $\Delta$ representing the largest distance between pair of the
points that were initially in the $\epsilon$-square box but have
spread out for an interval of time $t$.  $L^i_2(t)$ measures how an
area enclosing points grows. In practice, it is calculated by
$L^i_2(t)=\frac{A}{\epsilon^2}$, with $\epsilon^2$ representing the
area occupied by points in an $\epsilon$-square box, and $A$ the area
occupied by these points after spreading out for a time interval $t$.
There are $\tilde{N}_C$ boxes occupied by points which are taken into
consideration in the calculation of $L_k^i(t)$.  An order-$k$
expansion rate, $e_k(t)$, measures on average how a hypercube of
dimension $k$ exponentially grows after an interval of time $t$.  So,
$e_1$ measures the largest growth rate of nearby points, a quantity
closely related to the largest finite-time Lyapunov exponent
\cite{celso1994}.  And $e_2$ measures how an area enclosing points
grows, a quantity closely related to the sum of the two largest
positive Lyapunov exponents. In terms of expansion rates, Eqs.
(\ref{T}) and (\ref{I_C}) read $T = \frac{1}{e_1}\log{\left[
    \frac{1}{\epsilon} \right]}$ and $I_C = {e_1}(2 - D)$,
respectively, and Eqs. (\ref{d}) and (\ref{Icl}) read
$D(t)=\frac{e_2(t)}{e_1(t)}$ and $I_C^l=e_1(2-\tilde{D}_0)$,
respectively.

From the way we have defined expansion rates, we expect that $e_k \leq
\sum_{i=1}^k \lambda_i$.  Because of the finite time interval and the
finite size of the regions of points considered, regions of points
that present large derivatives, contributing largely to the Lyapunov
exponents, contribute less to the expansion rates. If a system has
constant derivative (hyperbolic) and has constant natural measure,
then $e_k=\sum_{i=1}^k \lambda_i$.

There are many reasons for using expansion rates in the way we have
defined them in order to calculate bounds for the MIR.  Firstly,
because they can be easily experimentally estimated whereas Lyapunov
exponents demand huge computational efforts.  Secondly, because of the
macroscopic nature of the expansion rates, they might be more
appropriate to treat data coming from complex systems that contains
large amounts of noise, data that have points that are not
(arbitrarily) close as formally required for a proper calculation of
the Lyapunov exponents.  Thirdly, expansion rates can be well defined
for data sets containing very few data points: the fewer points a data
set contains, the larger the regions of size $\epsilon$ need to be and
the shorter the time $T$ is.  Finally, expansion rates are defined in
a similar way to finite-time Lyapunov exponents and thus some
algorithms used to calculate Lyapunov exponents can be used to
calculate our defined expansion rates.

\section{Applications}

\subsection{MIR and its bounds in two coupled chaotic maps}

To illustrate the use of our bounds, we consider the following two
bidirectionally coupled maps{\small
\begin{eqnarray}
  X^{(1)}_{n+1}&=&2X^{(1)}_n+ \rho X^{(1)^{2}}_n + \sigma (X^{(2)}_n-X^{(1)}_n), \mbox{mod 1} \nonumber \\
  X^{(2)}_{n+1}&=&2X^{(2)}_n+ \rho X^{(2)^{2}}_n + \sigma (X^{(1)}_n-X^{(2)}_n), \mbox{mod 1}
\label{network_maps}
\end{eqnarray}}
\noindent
where $X_n^{(i)} \in [0,1]$. If $\rho=0$, the map is piecewise-linear
and quadratic, otherwise.  We are interested in measuring the exchange
of information between $X^{(1)}$ and $X^{(2)}$.  The space $\Omega$
is a square of sides 1. The Lyapunov exponents measured in the
space $\Omega$ are the Lyapunov exponents of the set
$\Sigma_{\Omega}$ that is the chaotic attractor generated by Eqs.
(\ref{network_maps}).

\begin{figure}[!h]
\centerline{\hbox{\psfig{file=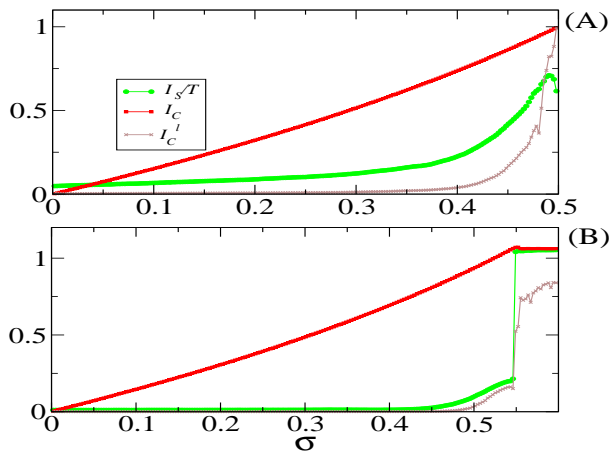,height=6cm,width=8cm,
viewport=0 10 450 470}}}
\caption{[Color online] Results for two coupled maps. $I_S/T$ [Eq.
  (\ref{MIR})] as (green online) filled circles, $I_C$ [Eq.
  (\ref{I_C})] as the (red online) thick line, and $I_C^l$ [Eq.
  (\ref{Icl})] as the (brown online) crosses.  In (A) $\rho=0$ and in
  (B) $\rho=0.1$.  The units of $I_S/T$, $I_C$, and $I_C^l$ are
  [bits/iteration].}
\label{figure4}
\end{figure}

The quantities $I_S/T$, $I_C$, and $I_C^l$ are shown in Fig.
\ref{figure4} as we vary $\sigma$ for $\rho=0$ (A) and $\rho=0.1$ (B).
We calculate $I_S$ using in Eq.  (\ref{IS}) the probabilities $P(i,j)$
in which points from a trajectory composed of $2,000,000$ samples fall
in boxes of sides $\epsilon$=1/500 and the probabilities $P(i)$ and
$P(j)$ that the points visit the intervals $[(i-1)\epsilon,i\epsilon[$
of the variable $X_n^{(1)}$ or $[(j-1)\epsilon,j\epsilon[$ of the
variable $X_n^{(2)}$, respectively, for $i,j=1,\ldots,N$.  When
computing $I_S/T$, the quantity $T$ was estimated by Eq.  (\ref{T}).
Indeed for most values of $\sigma$, $I_C \geq I_S/T$ and $I_C^l \leq
I_S/T$.

For $\sigma=0$ there is no coupling, and therefore the two maps are
independent from each other. There is no information being exchanged.
In fact, $I_C=0$ and $I_C^l \cong 0$ in both figures, since
$D=\tilde{D}_0=2$, meaning that the attractor $\Sigma_{\Omega}$ fully
occupies the space $\Omega$. This is a remarkable property of our
bounds: to identify that there is no information being exchanged when
the two maps are independent.  Complete synchronisation is achieved
and $I_C$ is maximal, for $\sigma>0.5$ (A) and for $\sigma \geq 0.55$
(B). A consequence of the fact that $D= \tilde{D}_0=1$, and therefore,
$I_C=I_C^l=\lambda_1$. The reason is because for this situation this
coupled system is simply the shift map, a map with constant natural
measure; therefore $P(i)=P(j)$ and $P(i,j)$ are constant for all $i$
and $j$. As usually happens when one estimates the mutual information
by partitioning the phase space with a grid having a finite resolution
and data sets possessing a finite number of points, $I_S$ is typically
larger than zero, even when there is no information being exchanged
($\sigma=0$).  Even when there is complete synchronisation, we find
non-zero off-diagonal terms in the matrix for the joint probabilities
causing $I_S$ to be smaller than it should be. Due to numerical
errors, $X^{(1)} \cong X^{(2)}$, and points that should be occupying
boxes with two corners exactly along a diagonal line in the subspace
$\Omega$ end up occupying boxes located off-diagonal and that have at
least three corners off-diagonal. The estimation of the lower bound
$I_C^l$ suffers from the same problems.
 
Our upper bound $I_C$ is calculated assuming that there is a
fictitious dynamics expanding points (and producing probabilities) not
only exponentially fast but also uniformly. The ``experimental''
numerical points from Eqs. (\ref{network_maps}) expand exponentially
fast, but not uniformly. Most of the time the trajectory remains in 4
points: (0,0), (1,1), (1,0), (0,1). That is the main reason of why
$I_C$ is much larger than the estimated real value of the $MIR$, for
some coupling strengths. If a two nodes in a dynamical network, such
as two neurons in a brain, behave in the same way the fictitious
dynamics does, these nodes would be able to exchange the largest
possible amount of information.

We would like to point out that one of the main advantages of
calculating upper bounds for the MIR ($I_S/T$) using Eq.  (\ref{I_C})
instead of actually calculating $I_S/T$ is that we can reproduce the
curves for $I_C$ using much less number of points (1000 points) than
the ones ($2,000,000$) used to calculate the curve for $I_S/T$. If
$\rho=0$, $I_C=-\ln{(1-\sigma)}$ can be calculated since
$\lambda_1=\ln{(2)}$ and $\lambda_2=\ln{(2-2\sigma)}$.

\subsection{MIR and its bounds in experimental networks of  Double-Scroll circuits}

We illustrate our approach for the treatment of data sets using a
network formed by an inductorless version of the Double-Scroll circuit
\cite{inductorless_chua}. We consider four networks of bidirectionally
diffusively coupled circuits.  Topology I represents two
bidirectionally coupled circuits, Topology II, three circuits coupled
in an open-ended array, Topology III, four circuits coupled in an
open-ended array, and Topology IV, coupled in an closed array. We
choose two circuits in the different networks (one connection apart)
and collect from each circuit a time-series of 79980 points, with a
sampling rate of $\delta=80.000$ samples/s.  The measured variable is
the voltage across one of the circuit capacitors, which is normalised
in order to make the space $\Omega$ to be a square of sides 1. Such
normalisation does not alter the quantities that we calculate. The
following results provide the exchange of information between these
two chosen circuits.  The values of $\epsilon$ and $t$ used to
course-grain the space $\Omega$ and to calculate $e_2$ in Eq.
(\ref{define_exp_rates}) are the ones that minimises
$|N_C(T,e_2)-\tilde{N}_C(\epsilon)|$ and at the same time satisfy
$N_C(T,e_2) \geq \tilde{N}_C(\epsilon)$, where
$N_C(T,e_2)=\exp^{Te_2(t)}$ represents the number of fictitious boxes
covering the set $\Sigma_{\Omega}$ in a compact fashion, when $t=T$.
This optimisation excludes some non-significant points that make the
expansion rate of fictitious points to be much larger than it should
be. In other words, we require that $e_2$ describes well the way most
of the points spread. We consider that $t$ used to calculate $e_k$ in
Eq.  (\ref{define_exp_rates}) is the time for points initially in an
$\epsilon$-side box to spread to 0.8$L$.  That guarantee that nearby
points in $\Sigma_{\Omega}$ are expanding in both directions within
the time interval $[0,T]$. Using $0.4L<t<0.8L$ produces already
similar results. If $t>0.8L$, the set $\Sigma_{\Omega}$ might not be
only expanding. $T$ might be overestimated.

\begin{figure}[t]
\includegraphics[height=7.0cm,width=7.0cm]
{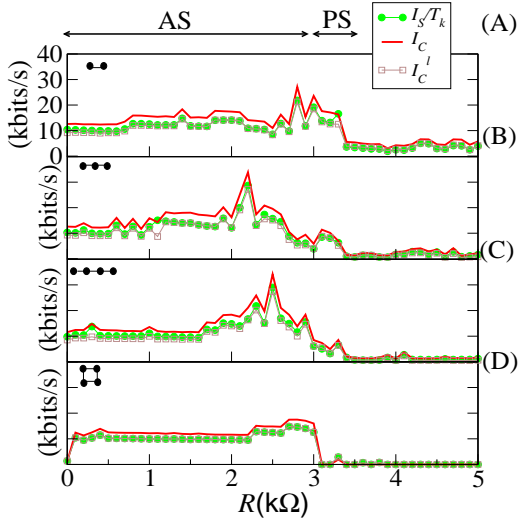}
\caption{[Color online] Results for experimental networks of
  Double-Scroll circuits. On the left-side upper corner pictograms
  represent how the circuits (filled circles) are bidirectionally
  coupled. $I_S/T_k$ as (green online) filled circles, $I_C$ as the
  (red online) thick line, and $I_C^l$ as the (brown online) squares,
  for a varying coupling resistance $R$.  The unit of these quantities
  shown in these figures is (kbits/s).  (A) Topology I, (B) Topology
  II, (C) topology III, and (D) Topology IV. In all figures,
  $\tilde{D}_0$ increases smoothly from 1.25 to 1.95 as $R$ varies
  from 0.1k$\Omega$ to 5k$\Omega$. The line on the top of the figure
  represents the interval of resistance values responsible to induce
  almost synchronisation (AS) and phase synchronisation (PS).}
\label{figure4_letter00}
\end{figure}

$I_S$ has been estimated by the method in Ref.  \cite{kraskov}.  Since
we assume that the space $\Omega$ where mutual information is being
measured is 2D, we will compare our results by considering in the
method of Ref.  \cite{kraskov} a 2D space formed by the two collected
scalar signals. In the method of Ref.  \cite{kraskov} the phase space
is partitioned in regions that contain 30 points of the continuous
trajectory. Since that these regions do not have equal areas (as it is
done to calculate $I_C$ and $I_C^l$), in order to estimate $T$ we need
to imagine a box of sides $\epsilon_k$, such that its area
$\epsilon_k^2$ contains in average 30 points. The area occupied by the
set $\Sigma_{\Omega}$ is approximately given by $\epsilon^2
\tilde{N}_C$, where $\tilde{N}_C$ is the number of occupied boxes.
Assuming that the 79980 experimental data points occupy the space
$\Omega$ uniformly, then on average 30 points would occupy an area of
$\frac{30}{79980}\epsilon^2 \tilde{N}_C$. The square root of this area
is the side of the imaginary box that would occupy 30 points. So,
$\epsilon_k=\sqrt{\frac{30}{79980} \tilde{N}_C}\epsilon$.  Then, in
the following, the ``exact'' value of the MIR will be considered to be
given by $I_S/T_k$, where $T_k$ is estimated by $T_k =
-\frac{1}{e_1}\log{(\epsilon_k)}$.

The three main characteristics of the curves for the quantities
$I_S/T_k$, $I_C$, and $I_C^l$ (appearing in Fig.
\ref{figure4_letter00}) with respect to the coupling strength are that
(i) as the coupling resistance becomes smaller, the coupling strength
connecting the circuits becomes larger, and the level of
synchronisation increases followed by an increase in $I_S/T_k$, $I_C$,
and $I_C^l$, (ii) all curves are close, (iii) and as expected, for
most of the resistance values, $I_C>I_S/T_k$ and $I_C^l \leq I_S/T_k$.
The two main synchronous phenomena appearing in these networks are
almost synchronisation (AS) \cite{femat_PLA1999}, when the circuits
are almost completely synchronous, and phase synchronisation (PS)
\cite{juergen_book}. For the circuits considered in Fig.
\ref{figure4_letter00}, AS appears for the interval $R \in [0,3]$ and
PS appears for the interval $R \in [3,3.5]$. Within this region of
resistance values the exchange of information between the circuits
becomes large. PS was detected by using the technique from Refs.
\cite{baptista_PHYSICAD2006,pereira_PRE2007}.

\subsection{MIR and its upper bound in stochastic systems}

\begin{figure}[!h]
\centerline{\hbox{\psfig{file=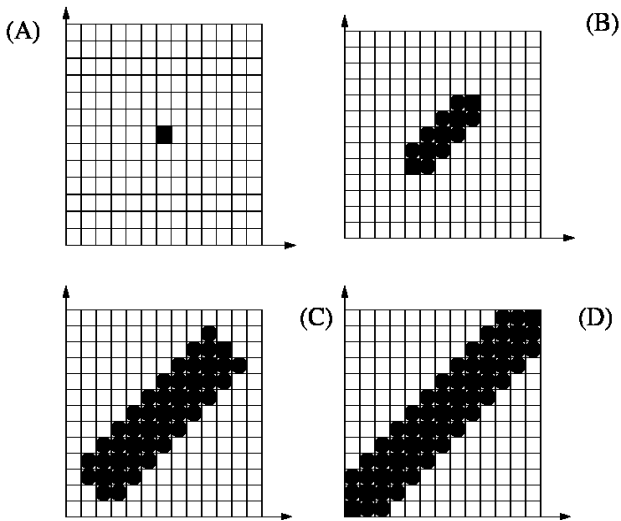,height=8cm,width=8cm,
viewport=0 -30 720 650}}}
\caption{(A) A small box representing a set of initial conditions.
  After one iteration of the system, the points that leave the
  initial box in (A) go to 4 boxes along the diagonal line [filled
  squares in (B)] and 8 boxes off-diagonal (along the transverse
  direction) [filled circles in (B)].  At the second iteration, the
  points occupy other neighbouring boxes as illustrated in (C) and
  after an interval of time $n=T$ the points do not spread any longer (D).}
\label{toy_model}
\end{figure}

To analytically demonstrate that the quantities $I_C$ and $I_S/T$ 
can be well calculated in stochastic systems, we
consider the following stochastic dynamical toy model illustrated in
Fig.  \ref{toy_model}. In it points within a small box of sides
$\epsilon$ (represented by the filled square in Fig.
\ref{toy_model}(A)) located in the centre of the subspace $\Omega$ are
mapped after one iteration of the dynamics to 12 other neighbouring
boxes.  Some points remain in the initial box.  The points that leave
the initial box go to 4 boxes along the diagonal line and 8 boxes
off-diagonal along the transverse direction. Boxes along the diagonal
are represented by the filled squares in Fig.  \ref{toy_model}(B) and
off-diagonal boxes by filled circles. At the second iteration, the
points occupy other neighbouring boxes, as illustrated in Fig.
\ref{toy_model}(C), and at the time $n=T$ the points do not spread any
longer, but are somehow reinjected inside the region of the attractor.
We consider that this system is completely stochastic, in the sense
that no one can precisely determine the location of where an initial
condition will be mapped. The only information is that points inside a
smaller region are mapped to a larger region.

At the iteration $n$, there will be $N_{d}=2^{1+n}+1$ boxes occupied
along the diagonal (filled squares in Fig. \ref{toy_model}) and
$N_{t}=2nN_d-C(\tilde{n})$ (filled circles in Fig. \ref{toy_model})
boxes occupied off-diagonal (along the transverse direction), where
$C(\tilde{n})=0$ for $\tilde{n}$=0, and $C(\tilde{n})>0$ for
$\tilde{n} \geq 1$ and $\tilde{n}=n-T-\alpha$. $\alpha$ is a small
number of iterations representing the time difference between the time
$T$ for the points in the diagonal to reach the boundary of the space
$\Omega$ and the time for the points in the off-diagonal to reach this
boundary.  The border effect can be ignored when the expansion along
the diagonal direction is much faster than along the transverse
direction.

At the iteration $n$, there will be
$N_C=2^{1+n}+1+(2^{1+n}+1)2n-C(\tilde{n})$ boxes occupied by points.
In the following calculations we consider that $N_C \cong
2^{1+n}(1+2n)$. We assume that the subspace $\Omega$ is a square whose
sides have length 1, and that $\Sigma \in \Omega$, so $L=\sqrt{2}$.
For $n>T$, the attractor does not grow any longer along the
off-diagonal direction. The time $n=T$, for the points to spread over
the attractor $\Sigma$, can be calculated by the time it takes for
points to visit all the boxes along the diagonal. Thus, we need to
satisfy $N_{d}\epsilon\sqrt{2}=\sqrt{2}$.  Ignoring the 1 appearing in
the expression for $N_{d}$ due to the initial box in the estimation
for the value of $T$, we arrive that $T >
\frac{\log{(1/\epsilon)}}{\log{(2)}}-1$. This stochastic system is
discrete. In order to take into consideration the initial box in the
calculation of $T$, we pick the first integer that is larger than
$\frac{\log{(1/\epsilon)}}{\log{(2)}}-1$, leading $T$ to be the
largest integer that satisfies
\begin{equation}
T < -\frac{\log{(\epsilon)}}{\log{(2)}}.
\label{TM_T}
\end{equation}

The largest Lyapunov exponent or the order-1 expansion rate of this
stochastic toy model can be calculated by
$N_{d}(n)\exp^{\lambda_1}=N_{d}(n+1)$, which take us to
\begin{equation}
\lambda_1=\log{(2)}.
\label{TM_lambda1}
\end{equation}
\noindent
Therefore, Eq. (\ref{TM_T}) can be rewritten as
$T=-\frac{\log{(\epsilon)}}{\lambda_1}$.

The quantity $D$ can be calculated by
$D=\frac{\log{(N_C)}}{\log{(N)}}$, with $n=T$. Neglecting $C(\tilde{n})$ and
the 1 appearing in $N_C$ due to the initial box, we have that $N_C
\cong 2^{1+T}[1+2^T]$. Substituting in the definition of $D$, we obtain 
$D = \frac{(1+T)\log{(2)} + \log{(1+2^T)}} {-\log{(\epsilon)}}$.
Using $T$ from Eq.  (\ref{TM_T}), we arrive at
\begin{equation}
 D = 1+r,
\label{TM_D}
\end{equation}
\noindent
where 
\begin{equation}
r=-\frac{\log{(2)}}{\log{(\epsilon)}} - \frac{\log{(1+2^T)}}{\log{(\epsilon)}}
\label{r}
\end{equation}
\noindent
Placing $D$ and $\lambda_1$ in $I_C=\lambda_1(2-D)$, give us
\begin{equation}
I_C = \log{(2)}(1-r).
\label{TM_IC}
\end{equation}

Let us now calculate $I_S/T$. Ignoring the border effect, and
assuming that the expansion of points is uniform, then 
$P(i,j)=1/N_C$ and $P(i)=P(j)=1/N=\epsilon$. At the iteration $n=T$,
we have that $I_S=-2\log{(\epsilon)}-\log{(N_C)}$. Since $N_C \cong
2^{1+T}[1+2^T]$, we can write that $I_S=-2\log{(\epsilon)} -
(1+T)\log{(2)} - \log{(1+2^T)}$. Placing $T$ from Eq.  (\ref{TM_T})
into $I_S$ takes us to $I_S=-\log{(2)} -\log{(\epsilon)}
-\log{(1+2^T)}$.  Finally, dividing $I_S$ by $T$, we arrive that
\begin{eqnarray}
\frac{I_S}{T} &=& \log{(2)} \left[1+\frac{\log{(2)}}{\log{(\epsilon)}} + \frac{\log{(1+2^T)}}{\log{(\epsilon)}}\right] \nonumber \\ 
    &=&
    \log{(2)}(1-r).
\label{TM_IST}
\end{eqnarray}
\noindent
As expected from the way we have constructed this model, Eq.
(\ref{TM_IST}) and (\ref{TM_IC}) are equal and $I_C = \frac{I_S}{T}$.

Had we included the border effect in the calculation of $I_C$, denote
the value by $I_C^b$, we would have typically obtained that $I_C^b
\geq I_C$, since $\lambda_2$ calculated considering a finite space
$\Omega$ would be either smaller or equal than the value obtained 
by neglecting the border effect. 
Had we included the border effect in the calculation
of $I_S/T$, denote the value by $I_S^b/T$, typically we would expect
that the probabilities $P(i,j)$ would not be constant. That is because the
points that leave the subspace $\Omega$ would be randomly reinjected back
to $\Omega$. We would conclude that $I_S^b/T \leq I_S/T$.  Therefore,
had we included the border effect, we would have obtained that $I_C^b
\geq I_S^b/T$.

The way we have constructed this stochastic toy model results in $D
\cong 1$. This is because the spreading of points along the diagonal
direction is much faster than the spreading of points along the
off-diagonal transverse direction. In other words, the second largest
Lyapunov exponent, $\lambda_2$, is close to zero. Stochastic toy
models which produce larger $\lambda_2$, one could consider that the
spreading along the transverse direction is given by
$N_{t}=N_d2^{\alpha n}-C(\tilde{n})$, with $\alpha \in [0,1]$.

\subsection{Expansion rates for noisy data with few sampling
  points}\label{expansion_rate}

In terms of the order-1 expansion rate, $e_1$, our quantities read
$I_C = {e_1}(2 - D)$, $T = \frac{1}{e_1}\log{\left[ \frac{1}{\epsilon}
  \right]}$, and $I_C^l=e_1(2-\tilde{D}_0)$. In order to show that our
expansion rate can be used to calculate these quantities, we consider
that the experimental system is uni-dimensional and has a constant
probability measure. Additive noise is assumed to be bounded with
maximal amplitude $\eta$, and having constant density.

Our order-1 expansion rate is defined as 
\begin{equation}
  e_1(t)=1/\tilde{N}_C \sum_{i=1}^{\tilde{N}_C} \frac{1}{t}  \log{[L_1^i(t)]}. 
\label{define_exp_rates_sup}
\end{equation}
\noindent
where $L_1^i(t)$ measures the largest growth rate of nearby points.
Since all it matters is the largest distance between points, it can be
estimated even when the experimental data set has very few data
points. Since, in this example, we consider that the experimental noisy
points have constant uniform probability distribution, $e_1(t)$ can be
calculated by
\begin{equation}
  e_1(t) = \frac{1}{t}  \log{\left[ \frac{\Delta+2\eta}{\delta+2\eta} \right]}. 
\label{define_exp_rates_sup1}
\end{equation}
\noindent
where $\delta+2\eta$ represents the largest distance between pair of
experimental noisy points in an $\epsilon$-square box and
$\Delta+2\eta$ represents the largest distance between pair of the
points that were initially in the $\epsilon$-square box but have
spread out for an interval of time $t$. The experimental system
(without noise) is responsible to make points that are at most
$\delta$ apart from each other to spread to at most to $\Delta$ apart
from each other. This points spread out exponentially fast according
to the largest positive Lyapunov exponent $\lambda_1$ by
\begin{equation}
\Delta=\delta \exp^{\lambda_1t}. 
\label{delta}
\end{equation}

Substituting Eq. (\ref{delta}) in (\ref{define_exp_rates_sup1}), and
expanding $\log$ to first order, we obtain that $e_1=\lambda_1$, and
therefore, our expansion rate can be used to estimate Lyapunov
exponents.

\section{Supplementary Information}

\subsection{Decay of correlation and First Poincar\'e Returns}

As rigorously shown in \cite{young}, the decay with time of the
correlation, $C(t)$, is proportional to the decay with time of the
density of the first Poincar\'e recurrences, $\rho(t,\epsilon)$, which
measures the probability with which a trajectory returns to an
$\epsilon$-interval after $t$ iterations.  Therefore, if
$\rho(t,\epsilon)$ decays with $t$, for example exponentially fast,
$C(t)$ will decay with $t$ exponentially fast, as well. The
relationship between $C(t)$ and $\rho(t)$ can be simply understood in
chaotic systems with one expanding direction (one positive Lyapunov
exponent). As shown in \cite{baptista_CHAOS2009}, the ``local'' decay
of correlation (measured in the $\epsilon$-interval) is given by
$C(t,\epsilon) \leq \mu(\epsilon) \rho(t,\epsilon) - \mu(\epsilon)^2$,
where $\mu(\epsilon)$ is the probability measure of a chaotic
trajectory to visit the $\epsilon$-interval.  Consider the shift map
$x_{n+1}=2x_n, \mbox{mod 1}$. For this map, $\mu(\epsilon)=\epsilon$
and there are an infinite number of possible intervals that makes
$C(t,\epsilon)=0$, for a finite $t$.  These intervals are the cells of
a Markov partition. As recently demonstrated by [P. Pinto, I.
Labouriau, M. S. Baptista], in piecewise-linear systems as the shift
map, if $\epsilon$ is a cell in an order-$t$ Markov partition and
$\rho(t,\epsilon)>0$, then $\rho(t,\epsilon)=2^{-t}$ and by the way a
Markov partition is constructed we have that $\epsilon=2^{-t}$. Since
that $\epsilon=\mu(\epsilon)=2^{-t}$, we arrive at that $C(t,\epsilon)
\leq 0$, for a special finite time $t$. Notice that $\epsilon=2^{-t}$
can be rewritten as $-\ln{(\epsilon)}=t\ln{(2)}$.  Since for this map,
the largest Lyapunov exponent is equal to $\lambda_1=\ln{(2)}$, then
$t=-\frac{1}{\lambda_1}\ln{(\epsilon)}$, which is exactly equal to the
quantity $T$, the time interval responsible to make the system to lose
its memory from the initial condition and that can be calculated by
the time that makes points inside an initial $\epsilon$-interval to
spread over the whole phase space, in this case $[0,1]$.

\subsection{$I_C$, and $I_C^l$ in larger networks and higher-dimensional
  subspaces $\Sigma_{\Omega}$}\label{higher}

Imagine a network formed by $K$ coupled oscillators. Uncoupled, each
oscillator possesses a certain amount of positive Lyapunov exponents,
one zero, and the others are negative. Each oscillator has dimension
$d$.  Assume that the only information available from the network are
two $Q$ dimensional measurements, or a scalar signal that is
reconstructed to a $Q$-dimensional embedding space. So, the subspace
$\Sigma_{\Omega}$ has dimension $2Q$, and each subspace of a node (or
group of nodes) has dimension $Q$. To be consistent with our previous
equations, we assume that we measure $M_{\Omega}=2Q$ positive Lyapunov
exponents on the projection $\Sigma_{\Omega}$. If $M_{\Omega} \neq
2Q$, then in the following equations $2Q$ should be replaced by
$M_{\Omega}$, naturally assuming that $M_{\Omega} \leq 2Q$.

In analogy with the derivation of $I_C$ and $I_C^l$ in a bidimensional
projection, we assume that if the spreading of initial conditions is
uniform in the subspace $\Omega$.  Then, $P(i)=\frac{1}{N^Q}$ represents the
probability of finding trajectory points in $Q$-dimensional space of
one node (or a group of nodes) and $P(i,j)=\frac{1}{N_C}$ represents
the probabilities of finding trajectory points in the $2Q$-dimensional
composed subspace constructed by two nodes (or two groups of nodes) in
the subspace $\Omega$. Additionally, we consider that the hypothetical
number of occupied boxes $N_C$ will be given by
$N_C(T)=\exp^{T(\sum_{i=1}^{2Q}\lambda_i)}$. Then, 
we have that $T=1/\lambda_1 \log{(N)}$, 
which lead us to
\begin{equation}
  I_C = \lambda_1(2Q-D).
\label{IC_HD}
\end{equation}
\noindent
Similarly to the way we have derived $I_C^l$ in a bidimensional
projection, if $\Sigma_{\Omega}$ has more than 2 positive Lyapunov
exponents, then
\begin{equation}
  I_C^l = \lambda_1(2Q-\tilde{D}_0).  
\label{IC_HD0}
\end{equation}

To write Eq. (\ref{IC_HD}) in terms of the positive Lyapunov
exponents, we first extend the calculation of the quantity $D$ to
higher-dimensional subspaces that have dimensionality 2Q,
\begin{equation}
D=1+\sum_{i=2}^{2Q}\frac{\lambda_i}{\lambda_1}, 
\label{DK}
\end{equation}
\noindent
\noindent
where $\lambda_1 \geq \lambda_2 \geq \lambda_3 \ldots \geq
\lambda_{2Q}$ are the Lyapunov exponents measured on the subspace
$\Omega$. To derive this equation we only consider that the
hypothetical number of occupied boxes $N_C$ is given by
$N_C(T)=\exp^{T(\sum_{i=2}^{2Q}\lambda_i)}$.

We then substitute $D$ as a function of these exponents (Eq.
(\ref{DK})) in Eq. (\ref{IC_HD}). We arrive at
\begin{equation}
I_C = (2Q-1)\lambda_1 - \sum_{i=2}^{2Q} \lambda_i.
\label{ICU_HD}
\end{equation}

\subsection{$I_C$ as a function of the positive Lyapunov exponents of
  the network}

Consider a network whose attractor $\Sigma$ possesses $M$ positive
Lyapunov exponents, denoted by $\tilde{\lambda}_i$, $i=1,\ldots,M$.
For a typical subspace $\Omega$, $\lambda_1$ measured on $\Omega$ is equal to
the largest Lyapunov exponent of the network. Just for the sake of
simplicity, assume that the nodes in the network are sufficiently well
connected so that in a typical measurement with a finite number of
observations this property holds, i.e., $\tilde{\lambda}_1 =
\lambda_1$.  But, if measurements provide that $\tilde{\lambda}_1 >>
\lambda_1$, the next arguments apply as well, if one replaces
$\tilde{\lambda}_1$ appearing in the further calculations by the
smallest Lyapunov exponent, say, $\tilde{\lambda_k}$, of the network
that is still larger than $\lambda_1$, and then, substitute
$\tilde{\lambda}_2$ by $\tilde{\lambda_{k+1}}$, and so on.  As before,
consider that $M_{\Omega}=2Q$.

Then, for an arbitrary subspace $\Omega$, $\sum_{i=2}^{2Q}
\lambda_i \leq \sum_{i=2}^{2Q} \tilde{\lambda}_i$, since a projection
cannot make the Lyapunov exponents larger, but only smaller or equal.

Defining 
\begin{equation}
\tilde{I}_C = (2Q-1)\lambda_1 - \sum_{i=2}^{2Q} \tilde{\lambda}_i. 
\label{icu_new}
\end{equation}
Since $\sum_{i=2}^{2Q} \lambda_i \leq \sum_{i=2}^{2Q}
\tilde{\lambda}_i$, it is easy to see that
\begin{equation}
\tilde{I}_C  \leq I_C.  
\label{icu_new1}
\end{equation}

So, $I_C$, measured on the subspace $\Sigma_{\Omega}$ and a function
of the $2Q$ largest positive Lyapunov exponents measured in
$\Sigma_{\Omega}$, is an upper bound for $\tilde{I}_C$, a quantity
defined by the $2Q$ largest positive Lyapunov exponents of the attractor
$\Sigma$ of the network.  Therefore, if the Lyapunov exponents of a
network are know, the quantity $\tilde{I}_C$ can be used as a way to
estimate how much is the MIR between two measurements of this network,
measurements that form the subspace $\Omega$.

Notice that $I_C$ depends on the projection chosen (the subspace
$\Omega$) and on its dimension, whereas $\tilde{I}_C$ depends on the
dimension of the subspace $\Sigma_{\Omega}$ (the number 2Q of positive
Lyapunov exponents). The same happens for the mutual information
between random variables that depend on the projection considered.

Equation (\ref{icu_new}) is important because it allows us to obtain an
estimation for the value of $I_C$ analytically. As an example, imagine
the following network of coupled maps with a constant Jacobian 

\begin{equation}
X^{(i)}_{n+1}=2X^{(i)}_n+ \sigma \sum_{j=1}^K {\mathbf A}_{ij}(X^{(j)}_n-X^{(i)}_n), \mbox{mod 1}, 
\label{network_maps_large}
\end{equation}
\noindent
where $X \in [0,1]$ and ${\mathbf A}$ represents the connecting adjacent matrix. If node $j$ connects to node $i$, then   ${\mathbf A}_{ij}=1$, and 0 otherwise. 

Assume that the nodes are connected all-to-all. Then, the $K$ positive
Lyapunov exponents of this network are: $\tilde{\lambda}_1=\log{(2)}$
and $\tilde{\lambda}_i=\log{2[1+\sigma]}$, with $i=2,K$. Assume also
that the subspace $\Omega$ has dimension $2Q$ and that $2Q$ positive
Lyapunov exponents are observed in this space and that
$\tilde{\lambda}_1 = \lambda_1$. Substituting these Lyapunov exponents
in Eq. (\ref{icu_new}), we arrive at
\begin{equation}
\tilde{I}_C = (2Q-1)\log{(1+\sigma)}.
\label{ICU_HD_ilustra}
\end{equation}
\noindent
We conclude that there are two ways for $\tilde{I}_C$ to increase.
Either one considers larger measurable subspaces $\Omega$ or one
increases the coupling between the nodes. This suggests that the
larger the coupling strength is the more information is exchanged
between groups of nodes.

For arbitrary topologies, one can also derive analytical formulas for
$\tilde{I}_C$ in this network, since $\tilde{\lambda}_i$ for $i>2$ can be
calculated from $\tilde{\lambda}_2$ \cite{baptista_PLA2010c}. One arrives at
\begin{equation}
\tilde{\lambda}_i(\omega_i \sigma/2)=\tilde{\lambda}_2(\sigma), 
\end{equation}
\noindent
where $\omega_i$ is the $i$th largest eigenvalue (in absolute value)
of the Laplacian matrix ${\mathbf L}_{ij}={\mathbf A}_{ij} + \mathbb{I} 
\sum_{j}{\mathbf A}_{ij}$.

\section{Conclusions}

Concluding, we have shown a procedure to calculate mutual information
rate (MIR) between two nodes (or groups of nodes) in dynamical
networks and data sets that are either mixing, or present fast decay
of correlations, or have sensitivity to initial conditions, and have
proposed significant upper ($I_C$) and lower ($I_C^l$) bounds for it,
in terms of the Lyapunov exponents, the expansion rates, and the
capacity dimension.  Since our upper bound is calculated from Lyapunov
exponents or expansion rates, it can be used to estimate the MIR
between data sets that have different sampling rates or experimental
resolution (e.g.  the rise of the ocean level and the average
temperature of the Earth), or between systems possessing a different
number of events. Additionally, Lyapunov exponents can be accurately
calculated even when data sets are corrupted by noise of large
amplitude (observational additive noise) \cite{mera,gao2006} or when
the system generating the data suffers from parameter alterations
(``experimental drift'') \cite{stefanski}.  Our bounds link
information (the MIR) and the dynamical behaviour of the system being
observed with synchronisation, since the more synchronous two nodes
are, the smaller $\lambda_2$ and $D_0$ will be. This link can be of
great help in establishing whether two nodes in a dynamical network or
in a complex system not only exchange information but also have linear
or non-linear interdependences, since the approaches to measure the
level of synchronisation between two systems are reasonably well known
and are been widely used.  If variables are synchronous in a time-lag
fashion \cite{juergen_book}, it was shown in Ref. \cite{blanc} that
the MIR is independent of the delay between the two processes. The
upper bound for the MIR could be calculated by measuring the Lyapunov
exponents of the network (see Supplementary Information), which are
also invariant to time-delays between the variables.

{\bf Acknowledgments} M. S. Baptista was partially supported by the
Northern Research Partnership (NRP) and Alexander von Humboldt
foundation. M. S. Baptista would like to thank A.  Politi for
discussions concerning Lyapunov exponents.  R.M. Rubinger, E.R.  V.
Junior and J.C.  Sartorelli thanks the Brazilian agencies CAPES, CNPq,
FAPEMIG, and FAPESP.

\end{document}